\documentclass[12pt]{article}
\usepackage{amssymb}

\usepackage[dvips]{epsfig}

\begin{document}

\author{\and A. de Souza Dutra\thanks{%
e-mail: dutra@feg.unesp.br}, Marcelo Hott\thanks{%
e-mail: hott@feg.unesp.br} \\
UNESP/Campus de Guaratinguet\'a-DFQ\\
Av. Dr. Ariberto Pereira da Cunha, 333\\
12516-410, Guaratinguet\'{a} - SP - Brazil \and C. A. S. Almeida \\
Departamento de F\'{\i}sica, Universidade Federal do Cear\'a\\
Centro de Ci\^{e}ncias, Campus do Pici, Caixa Postal 6030\\
60455-760, Fortaleza - CE - Brazil}
\title{Remarks on supersymmetry of quantum systems with position-dependent
effective masses}
\maketitle

\begin{abstract}
We apply the supersymmetry approach to one-dimensional quantum systems with
spatially-dependent mass, by including their ordering ambiguities
dependence. In this way we extend the results recently reported in the
literature. Furthermore, we point out a connection between these systems and
others with constant masses. This is done through convenient transformations
in the coordinates and wavefunctions. \vspace{1cm}

\end{abstract}

\thinspace

\newpage

Recently some of the authors of the present work did an analysis of the
classification of quantum systems with position-dependent mass regarding
their exact solvability \cite{Ambiguity}. On a similar basis Plastino
\textit{et al.} \cite{plastino}, applied the supersymmetric
quantum-mechanical approach to such systems, corresponding to effective
theories related to some solid state problems. In that paper the authors
considered the following kind of Shr\"{o}dinger equation
\begin{equation}
\left[ -\hslash ^{2}{\vec{\nabla}}\frac{1}{2\,m\left( \vec{r}\right) }{\vec{%
\nabla}}\,+\,V\left( \vec{r}\right) \right] \psi \left( \vec{r}\right)
\,=\,E\,\psi \left( \vec{r}\right) ,
\end{equation}

\noindent and succeed to show that some one-dimensional systems with
position-dependent effective mass have a supersymmetric partner system with
the same effective mass. They were also able to solve exactly some
particular cases by constructing the superpotential from the form of the
effective mass $m(x)$ and generalizing the concept of shape invariance.
However, this kind of physical problem is intrinsically ambiguous [1, 3-4],
and consequently the above Schr\"{o}dinger equation is a particular case of
the most general Hamiltonian, as originally proposed by von Roos \cite
{vonroos},

\begin{equation}
H_{VR}\,=\, -\frac{\hbar^2}{4}\left[ m^\delta (\vec r)\,{\vec \nabla}%
\,m^\kappa (\vec r)\,{\vec \nabla}\, m^\lambda (\vec r)\,+\, m^\lambda (\vec
r)\,{\vec \nabla}\,m^\kappa(\vec r)\,{\vec \nabla} \, m^\delta (\vec
r)\,\right] \,+ \, V(\vec r) ,
\end{equation}

\noindent whose classical limit is identical to the first one, and the
parameters are constrained by the condition $\delta +\kappa +\lambda =-1$.
Here we intend to extend the results of Plastino \textit{et al.}, in order
to accommodate this more general situation.

In fact, the problem of ordering ambiguity is a long standing one in quantum
mechanics. Some of the founders of quantum mechanics as Born, Jordan, Weyl,
Dirac and von Newmann worked on this matter, see for instance the excellent
critical review by Shewell \cite{shewell}. There are many examples of
physically important systems for which such an ambiguity is quite relevant.
For instance, we can cite the problem of impurities in crystals \cite
{luttinger, wannier, slater}, the dependence of nuclear forces on the
relative velocity of the two nucleons \cite{rojo, razavy}, and more recently
the study of semiconductor heterostructures \cite{bastard, weisbuch}. In
addition, taking into account the spatial variation of the semiconductor
type, some effective Hamiltonians were proposed with a position-dependent
mass for the carrier \cite{duke}-\cite{carlos}.

More recently, L\'{e}vy-Leblond \cite{leblond}, when discussing the case of
discontinuous masses, argued that there is a privileged ordering, namely the
one given in equation (1). This was achieved by choosing the continuity of
the wave function $\psi \left( x,t\right) $ and its derivative divided by
the mass $\frac{1}{m\left( x\right) }\frac{\partial \psi }{\partial x}.\,$
At this point it is interesting to quote a very recent work by Dekar et al
\cite{dekar}, where this ambiguity is still providently taken into account
from the beginning and then compared with the result of L\'{e}vy-Leblond.
Those authors call the attention to the fact that, even if one accept the
continuity condition used by L\'{e}vy-Leblond \cite{leblond}, there still
remains the question regarding the universality of this choice, as in the
case of continuous masses. As far as we know no one has proven such
universality. On the other hand, Morrow and Brownstein \cite{brownstein},
also addressing abrupt heterojunctions, had concluded that for the
Hamiltonian given in equation (2) the continuous quantities are
\begin{equation}
m\left( x\right) ^{\delta }\psi \,\left( x\right) \,\,\,\,\mathrm{and}%
\,\,\,\,\,m\left( x\right) ^{\left( \delta +\kappa \right) }\,\frac{\partial
\psi }{\partial x}\,\,,  \label{condition}
\end{equation}

\noindent where $2\,\delta +\,\kappa \,=\,-1$. This shows that there exists
at least a controversy about the possibility of removing the ambiguity, and
that seems to be intimately linked to the choice of the continuity
condition. Moreover, Henderson \textit{et al} \cite{henderson} suggested
some experiments, measuring the amplitude index of refraction, in order to
determine the ordering parameters ($\delta ,\,\kappa ,\lambda $) at the
discontinuities, once their values seem to be dependent of the type of
interfaces.

Finally, as it was seen in \cite{leblond} and \cite{Ambiguity}, it is
possible to include all the ambiguity into an effective potential and, since
the potentials play no role in the process of constructing a continuity
equation (conservation of the probability), it is easy to verify that the
conservation of the probability is absolutely unambiguous. Therefore, it is
at least doubtful that one could use any continuity condition to get rid of
the ambiguity. So it should be important to extend the supersymmetric
approach in such a way that these ambiguities could be taken into account.

The extension of the supersymmetric approach devised in \cite{plastino},
capable of generating those cases can be done by starting from the following
generalized ladder operators
\begin{eqnarray}
A\; &=&\;i\;a\;\left( m^{\alpha }\,\hat{p}\,m^{\beta }\,+\,b\,m^{\beta }\,%
\hat{p}\,m^{\alpha }\right) \,+\,\tilde{W}  \nonumber \\
A^{\dagger }\; &=&\;-\,i\;a\;\left( m^{\beta }\,p\,m^{\alpha
}\,+\,b\,m^{\alpha }\,p\,m^{\beta }\right) \,+\,\tilde{W},
\end{eqnarray}

\noindent where $\hat{p}=-i\hbar \frac{d}{dx}$ is the momentum operator
acting on all factors to the right, $\tilde{W}$ is the generalized quantum
superpotential and $\alpha $, $\beta $, $a$, $b$ are arbitrary parameters
related by $a=\frac{1}{\sqrt{2}\left( b+1\right) },\,\,\alpha +\beta \,=-\,%
\frac{1}{2}$.

The supersymmetric partner Hamiltonians $H_{1}\,=\,\,A^{\dagger }\,A$ and $%
H_{2}=\,A\,A^{\dagger }$\cite{cooper} can be written as
\begin{equation}
H_{i}=\,\frac{1}{2\,m}\left[ {\hat{p}}^{2}\,+\,i\,\hbar \left( \frac{%
m^{\prime }}{m}\right) \,{\hat{p}}\right] \,+V_{i}
\end{equation}

\noindent where $m^{\prime }=\frac{dm}{dx}$ , and $i=1,2$. One can verify
that the case studied by Plastino \textit{et al.} is recovered by choosing $%
\alpha =-\frac{1}{2}$ and $b=0$. In the above equations, we have defined the
superpartner potentials respectively as
\[
V_{1}\,=\,\tilde{W}^{2}-\frac{\hbar }{\sqrt{2m}\left( b+1\right) }\left[
\left( \beta -\alpha \right) \left( b-1\right) \left( \frac{m^{\prime }}{m}%
\right) \tilde{W}+\left( b+1\right) \tilde{W}^{\prime }\right]
\]
\begin{equation}
-\,\frac{\hbar ^{2}}{2m\left( b+1\right) ^{2}}\left( \beta +b\alpha \right)
\left\{ \left( b+1\right) \left( \frac{m^{\prime \prime }}{m}\right) +\left[
\beta \left( 2b+1\right) +\alpha \left( b+2\right) -\left( b+1\right)
\right] \left( \frac{m^{\prime }}{m}\right) ^{2}\right\} ,
\end{equation}

\noindent and
\[
V_{2}\,=\,\tilde{W}^{2}+\frac{\hbar }{\sqrt{2m}\left( b+1\right) }\left[
\left( \alpha -\beta \right) \left( b-1\right) \left( \frac{m^{\prime }}{m}%
\right) \tilde{W}+\left( b+1\right) \tilde{W}^{\prime }\right] +
\]
\begin{equation}
-\,\frac{\hbar ^{2}}{2m\left( b+1\right) ^{2}}\left( \alpha +b\beta \right)
\left\{ \left( b+1\right) \left( \frac{m^{\prime \prime }}{m}\right)
\,+\left[ \alpha \left( 2b+1\right) +\beta \left( b+2\right) -\left(
b+1\right) \right] \left( \frac{m^{\prime }}{m}\right) ^{2}\right\} .
\end{equation}
\noindent \qquad \qquad

Now, as an example, we apply the above results to the particular case of
potentials with the harmonic oscillator spectrum, similarly to which was
done in \cite{plastino}. The shape invariance is guaranteed by the condition
$V_{2}\left( x;k\right) \,=\,V_{1}\left( x;k\right) \,+\,k$, where $k$ is a
uniform energy shift. This lead us to the following equation obeyed by the
superpotential $\tilde{W}$,
\begin{equation}
\frac{\sqrt{2}\hbar }{\sqrt{m}}\,\tilde{W}^{\prime }\,+\,\frac{\hbar ^{2}}{2m%
}\left( \frac{b-1}{b+1}\right) \left( \frac{1}{2}\,+\,2\,\alpha \right)
\left[ \frac{m^{\prime \prime }}{m}\,+\,\frac{3}{2}\left( b-1\right) \left(
\frac{m^{\prime }}{m}\right) ^{2}\right] \,=\,k.  \label{W'}
\end{equation}

Note that the shape invariance as imposed implies that the partner
potentials differ only by a constant term $k$. So the expression of $V_{2}$
can be written through creation and annihilation operators as those written
in above. Consequently, one can repeat the above procedure again and again,
obtaining $V_{j+1}\left( x,a_{j}\right) \,=\,V_{j}\left( x,a_{j+1}\right)
+\,k$,

\noindent where $a_{j}$ stands for the potential and ambiguity parameters,
and the energy for the case of the harmonic oscillator type potential \cite
{plastino} is given by $E_{n}=n\,k$.

Finally, the corresponding eigenfunctions are obtained by successive
applications of creation operators, as in the usual supersymmetry procedure.
Another way to verify the possibility of constructing all the energy
eigenspectra from the above shape invariance imposition, is to note that $%
H_{2}$ can be cast into the form $H_{2}=A^{\dagger }A+k$.

\noindent So one can construct a Hamiltonian $H_{3}$, supersymmetric partner
of $H_{2}$ given by
\begin{equation}
H_{3}=A\,A^{\dagger }+k=\,\frac{1}{2\,m}\left[ {\hat{p}}^{2}\,+\,i\,\hbar
\left( \frac{m^{\prime }}{m}\right) \,{\hat{p}}\right] +V_{3},
\end{equation}

\noindent where $V_{3}=V_{2}+k=V_{1}+2\,k$. In its turn, it can be rewritten
as $H_{3}=A^{\dagger }A+2\,k$.

This can be done in successive steps, so that after $n$ repetitions one get
\begin{equation}
H_{n+1}=A^{\dagger }A+n\,k=\,\frac{1}{2\,m}\left[ {\hat{p}}^{2}\,+\,i\,\hbar
\left( \frac{m^{\prime }}{m}\right) \,{\hat{p}}\right]
+V_{1}+\,n\,k=H_{1}+n\,k.
\end{equation}

Following this procedure, it can be proved that the energy levels of $H_{2}$
are the same of $H_{1}$, except by its ground state. In doing so, one can
verify that $E_{n+l}^{\left( 1\right) }=E_{n+l-1}^{\left( 2\right)
}=E_{n+l-2}^{\left( 3\right) }=\cdot \cdot \cdot =E_{n}^{\left( l\right) }$,
which lead us to the harmonic oscillator type energy, as expressed in above.

The comparison of the generalized potential $V_{1}(x)$ with that obtained in
reference \cite{plastino} can be done if one substitutes

\[
m(x) = \left( \frac{\gamma + x^{2}}{1 + x^{2}} \right)^2 \,
\]

\noindent in equation(\ref{W'}). Then one obtains

\[
\tilde{W}(x)=\frac{kx}{\sqrt{2}}+\frac{\gamma -1}{\sqrt{2}}\ k\,\arctan {x}%
+2\left( \frac{b-1}{b+1}\right) \left( \frac{1}{2}+2\alpha \right) \left\{
\frac{\gamma -1}{\sqrt{2}}\frac{x}{(\gamma +x^{2})^{2}}+\right.
\]
$\ $%
\[
\left. +6b\left[ \frac{1}{8(\gamma -1)\gamma ^{3/2}}\ (3\gamma ^{2}+6\gamma
-1)\arctan \frac{x}{\sqrt{2}}-\frac{1}{\gamma -1}\arctan x+\right. \right.
\]

\begin{equation}
\left. \left. +\ \frac{2\gamma (\gamma -1)+(3\gamma ^{2}-6\gamma +7)(\alpha
+x^{2})}{4\gamma (\gamma +x^{2})}\right] \right\} .
\end{equation}

The final expression for the potential corresponding to the above
superpotential is very complicate, however one can obtain it
straightforwardly. Consequently, we present in fig. 1, two
particular choices of the ambiguity parameters. One of them
recovers the result appearing in \cite{plastino}. In fact, there
exist other qualitatively different situations which we are not
going to treat here, because it is out of the scope of this work.

\begin{figure}[tbp]
\begin{center}
\begin{minipage}{20\linewidth}
\epsfig{file=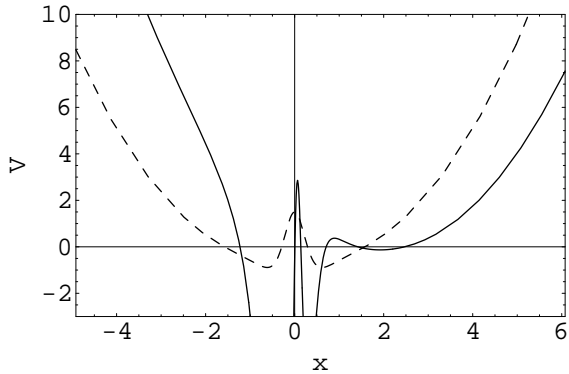}
\end{minipage}
\end{center}
\caption{Effective Potential for two particular values of the
ambiguity parameter $b $ and $\alpha = - 1/2$. The case of Pastino
et al: $b\, = \, 0 $ (dashed line) and another possible one $b\, =
\, 0.5 $ (solid line)} \label{fig:fig}
\end{figure}

\smallskip On devising one way to implement supersymmetry in quantum
mechanics for systems with position-dependent mass we have found how it is
possible in some particular cases. We show through a convenient
transformation of variables and redefinition of the wave function that
systems with position-dependent mass can be mapped into others with a
constant one, sharing the same spectrum and for which the quantum
supersymmetric approach is usually well known.

One-dimensional systems with position-dependent mass are in general
described by Hamiltonian (2) and consequently by the effective Schr\"odinger
equation

\begin{equation}
- \frac{\hbar^2}{2 \, m} \left( \frac{{d^2} \psi}{dx^2} - \frac{m^\prime}{m}
\frac{d \psi}{dx}\right) + V_{eff}(\delta ,\kappa ,\lambda ;x) \, \psi = E
\psi \, ,
\end{equation}

\noindent where

\begin{equation}
V_{eff}(\delta ,\kappa , \lambda ;x) = V(x) - \frac{\hbar^2}{4 \, m} \left[
(\delta + \lambda) \frac{m^{\prime \prime}}{m} - 2 (\delta + \lambda +
\delta\lambda) \left( \frac{m^\prime}{m} \right)^2 \right] \, .
\end{equation}

We note that the transformation of variable

\begin{equation}
u = \int^{x} \sqrt{2 \, m(z)} dz \, ,  \label{var}
\end{equation}

\noindent and the redefinition of the wave function

\[
\psi (u)=\left[ m(u)\right] ^{1/4}\varphi (u),
\]

\noindent leaves us with the following Schr\"{o}dinger equation, with mass
equals to unity.

\begin{equation}
-{\hbar ^{2}}\frac{d^{2}}{du^{2}}\varphi (u)+U_{eff}(\delta ,\lambda
;u)\,\varphi (u)=E\varphi (u)\,,  \label{constm}
\end{equation}

\noindent where

\begin{equation}
U_{eff}(\delta ,\lambda ;u)=V\left( u\right) -\frac{\hbar ^{2}}{4\,m}\left\{
\left[ 1+2(\delta +\lambda )\right] \frac{{d^{2}}m}{du^{2}}-\left( \frac{5}{4%
}+3\delta +3\lambda +4\delta \lambda \right) \left( \frac{dm}{du}\right)
^{2}\right\} \,.
\end{equation}

If one can implement supersymmetry for the above potential it can naturally
be done for the corresponding position-dependent mass system.

We recall that the potentials considered by Plastino \textit{et al.} can be
recovered by taking $\delta =\lambda =0$ and that is the case we consider
together with $U_{eff}(u)$ given by
\begin{equation}
U_{eff}(u)=\,\frac{\beta ^{2}}{4}\,u^{2}\,+\,\frac{\beta }{2},  \label{ueff1}
\end{equation}

\noindent and

\begin{equation}
U_{eff}(u)=C\,\Gamma \,e^{\Gamma u}+\left( B+C\,e^{\Gamma u}\right) ^{2}\,,
\label{ueff}
\end{equation}

\noindent respectively for the two examples considered by Plastino et al
\cite{plastino}. Moreover $\beta ,$ $B,C$ and $\Gamma $ are constants. The
first one corresponds to the harmonic oscillator and the second is the Morse
potential. The supersymmetric treatment for them is well known \cite{cooper}.

For these cases, equation (\ref{constm}) can be factorized as

\[
\mathcal{A}^{\dagger }\mathcal{A\ }\varphi (u)=E\,\varphi (u)\,,
\]

\noindent where

\[
\mathcal{A}=\frac{d}{du}+\frac{\beta }{2}\,u\,\,\, and \,\,\,\,\mathcal{A}=%
\frac{d}{du}+B+C\,e^{\Gamma u}\,,
\]

\noindent respectively. The action of this operator on the ground state $%
\left( \mathcal{A}\varphi _{0}(u)=0\right) $can be transformed
into\linebreak ${\tilde{A}}\,\psi _{0}(u)=0$, where

\begin{equation}
{\tilde{A}}=\frac{d}{du}-\frac{1}{4\,m}\frac{dm}{du}+\,\frac{\beta }{2}%
\,u\,\, and \,\,\,\,{\tilde{A}}=\frac{d}{du}-\frac{1}{4\,m}\frac{dm}{du}%
+B+C\,e^{\Gamma u}\,  \label{ann}
\end{equation}

\noindent respectively, and $\psi _{0}(u)=\left[ m(u)\right] ^{1/4}\varphi
_{0}(u)$.

On its turn the transformation of variable (\ref{var}) transforms the
operator (\ref{ann}) into the one prescribed by Plastino \textit{et al.} for
systems with harmonic oscillator (HO) and Morse-like (ML) spectra and
position-dependent mass, namely

\begin{equation}
A(x) = \frac{1}{\sqrt{2 \, m}} \frac{d}{dx} + W(x) \,
\end{equation}

\noindent where
\begin{equation}
W_{HO}(x)=\frac{1}{2}\,\frac{d}{dx}\left( \frac{1}{\sqrt{2\,m}}\right) +%
\frac{\beta }{2}\int^{x}\sqrt{2\left( z\right) }\,dz,
\end{equation}

\noindent and

\[
W_{ML}(x)=B+\frac{1}{2}\frac{d}{dx}\left( \frac{1}{\sqrt{2\,m}}\right)
+C\,e^{-\Gamma \int^{x}\sqrt{2\,m(z)}dz}\,.
\]

Then we have shown how systems with position-dependent mass can be mapped
into isospectral ones with constant mass for which the supersymmetric
approach can be implemented. This analysis is very important because in the
course of demonstration one can verify why the superpartner potentials $%
V_{1}(x)$ and $V_{2}(x)$ in equations (6) and (7) should depend on the
position-dependent mass in such a way that the corresponding Schr\"{o}dinger
equations could be exactly solvable, allowing one to understand to which
class of solvability it belongs \cite{Ambiguity}. Finally, it is important
to remark that such an analysis can easily be extended to the case where
some discontinuities are present. This can be achieved by performing the
transformation (\ref{var}) for each continuous interval under consideration
and imposing the continuity conditions ( \ref{condition}) at the
discontinuities.

\bigskip

\noindent \textbf{Acknowledgments:} The authors are grateful to CNPq and
FAPESP for partial financial support.

\end{document}